\documentclass[english]{article}
\usepackage[T1]{fontenc}
\usepackage[latin9]{inputenc}
\usepackage{color}
\usepackage{graphicx}



\usepackage{babel}
\begin{document}

\title{\textcolor{black}{Shiga Toxin Detection Methods : A Short Review}}


\author{Y. Castaño Guerrero, G. González-Aguilar$^{*}$}

\maketitle
$^{*}${\footnotesize{INESC-Porto UOSE Optoelectronics and Sensors
Unit, Faculty of Science University of Porto}}{\footnotesize \par}

\marginpar{\textit{Keywords}: Shiga-toxin, detection, biosensors, food.}

\begin{center}
------------------------------------------------------------------------------------------
\par\end{center}
\begin{abstract}
The Shiga toxins comprise a family of related protein toxins secreted
by certain types of bacteria. \textit{Shigella dysenteriae}, some
strain of \textit{Escherichia coli} and other bacterias can express
toxins which caused serious complication during the infection. Shiga
toxin and the closely related Shiga-like toxins represent a group
of very similar cytotoxins that may play an important role in diarrheal
disease and hemolytic-uremic syndrome. The outbreaks caused by this
toxin raised serious public health crisis and caused economic losses.
These toxins have the same biologic activities and according to recent
studies also share the same binding receptor, globotriosyl ceramide
(Gb3). Rapid detection of food contamination is therefore relevant
for the containment of food-borne pathogens. The conventional methods
to detect pathogens, such as microbiological and biochemical identification
are time-consuming and laborious. The immunological or nucleic acid-based
techniques require extensive sample preparation and are not amenable
to miniaturization for on-site detection. In the present are necessary
of techniques of rapid identification, simple and sensitive which
can be employed in the countryside with minimally-sophisticated instrumentation.
Biosensors have shown tremendous promise to overcome these limitations
and are being aggressively studied to provide rapid, reliable and
sensitive detection platforms for such applications. 
\end{abstract}
\begin{center}
------------------------------------------------------------------------------------------
\par\end{center}

\tableofcontents{}

\rule[0.1cc]{0.7\columnwidth}{1pt}

\section{Introduction}

\subsection{Preliminaries}

Shiga toxin was isolated from \textit{Shigella dysenteriae} serotype
1 by Kiyoshi Shiga in 1898, which named it \textit{Bacillus dysenterie}
\cite{lee_phylogenetic_2007}. In his work, Shiga described the production
of the toxic factors actually known as Shiga toxin \cite{trofa_dr._1999}.
Shiga toxin-producing\textit{ Escherichia coli} (STEC) was discovered
in 1977, and it was associated with the clinical hemolytic-uremic
syndrome (HUS) in 1983, a life-threatening condition characterized
by hemolytic anemia, thrombocytopenia, and renal failure \cite{thorpe_shiga_2004}.
This toxin has been also associated with hemorrhagic colitis and other
severe disease conditions. Most of the work in the identification
and characterization of Shiga toxin has been focused on\textit{ E.
coli }O157:H7 strains, although many cases of Shiga toxin associated
disease were caused by other serotypes of \textit{E. coli} \cite{lee_phylogenetic_2007}. 

The main reservoirs of STEC are cattle and sheep, but other animals
are recognized as a risk factor such as: deer, dogs, birds and horses.
Nevertheless, new vehicles of infection resulting from environmental
contamination and the intensive farming are continuously been identified
\cite{khan_antibiotic_2002}. Shiga toxin-producing \textit{Escherichia
coli} may cause diarrhea, bloody diarrhea and hemorrhagic colitis.
The transmission of STEC may occurs through contaminated foods, such
as ground beef, through contaminated water and by person-to-person
interaction \cite{hermos_shiga_2011}. Fecal-oral transmission is
also a common mode of transmission. Other ways to acquire that disease
is the direct contact with animals on farms or at zoos. The leafy
greens and unpasteurized apple cider are other recognized exposure
sources. The transmission from person-to-person can occur directly
(households, child care centers, institutions) or indirectly (contaminated
drinking or recreational water) \cite{Bentancor_2012}. The ingestion
of undercooked hamburgers in fast-food restaurants, was associate
with the production of hemorrhagic colitis (HC), produced by STEC
\cite{Lali_Growther_2011}. Stool cultures from patients infected
in these episode, yielded a previously rarely isolated\textit{ E.
coli} serotype, O157:H7. It has been reported that many \textit{E.
coli} strains isolated from these inpatients with diarrheal illness
produce a Shiga-like toxin (SLT), including one of the strains that
produce the Vero cytotoxin \cite{karmali_sporadic_1983}. Afterward
O\textquoteright{}Brien \emph{et al}., showed that Shiga-like toxin
and the Vero cytotoxin were the same toxin \cite{tesh_pathogenic_1991}.
Bacterial infections caused by Shiga toxin are responsible of many
disease and the death of a great number of people. These infections
have become a growing threat to the human health with worse manifestations
in developed countries where the bacterias able to release Shiga-like
toxins has been found in different food, including milk, apple juice
and vegetables. 

Surveillance data from the Center for Disease Control (CDC) of the
U.S. for the 1998 calendar year showed an increase in the number of
confirmed outbreaks of \textit{E. coli} O157:H7 infection, from an
average of 31 per year between 1994 and 1997 to 42 in 1998. In most
of the Asian countries, STEC is not yet a major health problem, except
in Japan, where 29 outbreaks were reported between 1991 to 1995 \cite{tesh_pathogenic_1991}.
The first documentation of outbreak was produced by an episode involving
strain O157:H7, in 1982 causing hemorrhagic colitis. Since then, the
incidence of this strain in the diseases has increased annually \cite{mauro_shiga_2011}.
An important massive outbreak of O157:H7 that gained particular attention
happened in the western U.S. in 1993, having 501 cases, 151 hospitalizations
and 45 cases of HUS. This outbreak was linked to consumption of undercooked
hamburger meat at a fast food restaurants \cite{hauswaldt_lessons_2013}.
More recently, the transmission of foodborne bacteria remains an important
public health threat because of the increased consumption of fresh
vegetables and fruits, and the increased consumption of foods in public
restaurants. The 80 \% of cases of bacterial diarrhea that occurs
every year in the United States are a result of foodborne transmission.
Shiga toxin\textendash{}producing\textit{ E. coli} is among the four
most commonly reported bacterial enteropathogens and contributes to
an estimated cost of \$7 billion annually. STEC in particular, causes
approximately 100,000 illnesses, 3,000 hospitalizations and 90 deaths
annually in the United States \cite{Clyde_Collins_2010}.

Numerous outbreaks, as well as sporadic cases of STEC infections and
HUS, have been documented worldwide. The largest outbreaks were recorded
in industrialized countries. The modern industrialized large-scale
food production might serve as a widespread vector in cases of food
contamination. In Germany, STEC were commonly recognized as pathogens
causing rare but severe disease almost exclusively in younger children.
Before 2011, about 1,000 STEC infections per year and fewer than 100
cases of HUS were registered. The 2011 outbreak of a STEC O104:H4
in Northern Germany started at the beginning of May, reaching its
peak on May 22. This an unusual outbreak of diarrhea with HUS affected
Germany and was spread to other European countries, United States
and Canada. Until July 4, when the end of the outbreak was officially
declared 4075 cases: 3935 in Germany, with 48 dead, while in set of
the other countries affected during this outbreak, were reported 140
cases \cite{cui_shiga_2013-1}. 

The most comparable historical outbreak to the recent German epidemic,
was took place in Japan in 1996 (Sakai city outbreak). About 6,000
persons were affected after white radish sprouts had been served at
school canteens, exposing 47,000 children to the contaminated food.
The incidence of HUS in the Sakai city outbreak was considerably lower
(106 cases out of 6,000 infections). Another outbreak of a STEC O157:H7
have occurred in 1996, in Scotland, involving 512 persons, of whom
279 cases were microbiologically confirmed. One of the largest non-O157
STEC outbreaks was caused in Oklahoma in 2008, which reported 341
cases with gastroenteritis, 71 patients required hospitalization and
1 patient died \cite{piercefield_ew_hemolytic_2010}. 

The virulence of E. coli 0157:H7 can partially be attributed to its
ability to establish infection at low doses in humans. The majority
of STEC infections present with hemorrhagic colitis as 91\% of patients
give a history of bloody diarrhea at some point during their illness.
Significant morbidity and mortality secondary to infection is attributed
to the development of HUS \cite{Clyde_Collins_2010}. Among the \textit{E.
coli} O157:H7 has received the most attention by the scientific and
regulatory community because of its association with several large
outbreaks of human illness with severe manifestations \cite{MICHAEL_2011}.
Even when the growing knowledge about toxins and its interaction with
cells have allowed the production of molecules that help to treat
this type of infection \cite{sandvig_shiga_2001}, this help some
times comes out of time, as most of the above infections where confirmed
several days after the initial episode. 

It is known that available methods are not characterized by its speed.
However, it is our intuition that biochemically-driven toxin detection
can improve the speed of diagnosis kits for bacterial diseases. Currently,
no extensive research on Stx identification and quantification is
available. The purpose of this work is to review the state-of-the
art techniques for Stx detection, trying to describe all the biological
and (bio)chemical approaches for its diagnostics.

\section{The Shiga family of toxins}

\subsection{Structural Characteristics}

Shiga Toxin and Shiga-like Toxin belong to large family of plants
and bacterial toxins. The real Shiga Toxin (Stx) is produced by \textit{Shigella
dysenteriae }and is almost identical to Shiga-like toxin 1 (Stx1).
Shiga-like toxin is secreted by some strains of \textit{Escherichia
coli }(STEC)\textit{, }and also\textit{ Citrobacter freundii, Aeromonas
hidrofilas, Aeromonas caviae} y \textit{Enterobacter cloacae} have
been reported able to express the toxin. Although the Shiga-like toxin
is have a similar structure, do not have the same effect on cells
\cite{paton_pathogenesis_1998} . 

As said before, the virulence factor of Shiga toxin, also known as
verotoxin, causes hemorrhagic colitis and diarrhea and in the most
severe cases leads to the lethal hemolytic-uremic syndrome. The primary
virulence factor in systemic host responses produced by clinical isolates
of STEC is Shiga toxin type 2 (Stx2), but some isolates produce both
Stx1 and Stx2, or more rarely only Stx1 \cite{zhang_quinolone_2000}. 

The two serological types of verotoxin, Stx1 and Stx2, are formed
by a active enzymatic complex with a subunit A and five subunit B
(Figure 1) \cite{van_hattum_functional_2012}. The fragment A has
an internal disulfide bond which when processed proteolytically generates
two subunit: A1 and A2. The A1 fragment inhibits protein synthesis
after it is released in the cytosol by the elimination of one adenine
from the 28S RNA of the 60S ribosomal subunit \cite{endo_site_1988}.
The B pentamer, of 89 aminoacids, can join the terminal of disaccharide
galabiose (gal-$\alpha$1,4-gal) in the surface of the host cells.
This interaction carbohydrate-toxin can be used to STEC detection
\cite{nagy_glycopolydiacetylene_2008}. 

\begin{figure}
\begin{centering}
\includegraphics[scale=0.7]{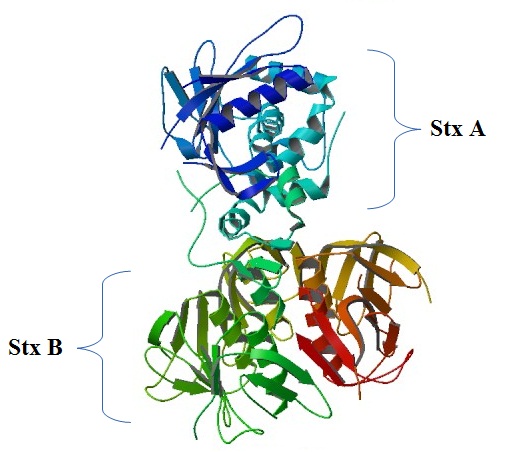}
\par\end{centering}

\caption{ 3-D structure of Shiga toxin A and B subunits . (Taken from RCSB
Protein Data Bank)}
\end{figure}

The molecular weight of the toxin is around 70kD, composed by subunit
A of 32 kD and each subunit B of 7.7 kD. Stx1 is virtually identical
to Stx, differing in only one aminoacid residue, whereas the Stx2
isoforms share less sequence similarity with Stx (60\%). Toxins Stx1
and Stx2 have similar structure but differ in their sequences (Figure
2): the fragment A has 315 aminoacid, while for Stx2 the subunit A
show 318 aminoacid \cite{sandvig_shiga_2001}. Although their primary
sequence of aminoacid are related, Stx1 and Stx2 are immunologically
different: both are able to join the Gb3 receptor but they do not
target same organs and tissues\cite{obrig_escherichia_2010}.\textcolor{black}{{}
Gb3 is expressed in many cells of the body human, but the fact that
amongst the more common complication hemorragic colitis and the HUS
are prevalent, suggests that the infection is directed to specific
organs. For example, Gb3 is expressed in the kidneys, mainly in the
pediatric glomeruli, and when the kidneys becomes adult this expression
decreases or is lost. This fact explains why HUS observed in children
is mainly caused by Stx1 \cite{rutjes_differential_2002}. It has
been reported that 90\% of HUS cases occurs in children younger than
three years old. Furthermore has been observed that Stx1 binds little
or nothing in adults renal glomeruli\cite{chark_differential_2004}. }

The affinity of Stx1 by the receptor Gb3 is ten times bigger than
with Stx2. The Shiga toxin genes are located in the bacteriophage
(bacteria virus), which is associated with all pathogenic STEC \cite{engedal_shiga_2011}.
The Stx1 and Stx2 are secreted by different ways and are translated
in different ways across the outer membrane \cite{shimizu_2007}. 

\begin{figure}
\begin{centering}
\includegraphics[scale=0.5]{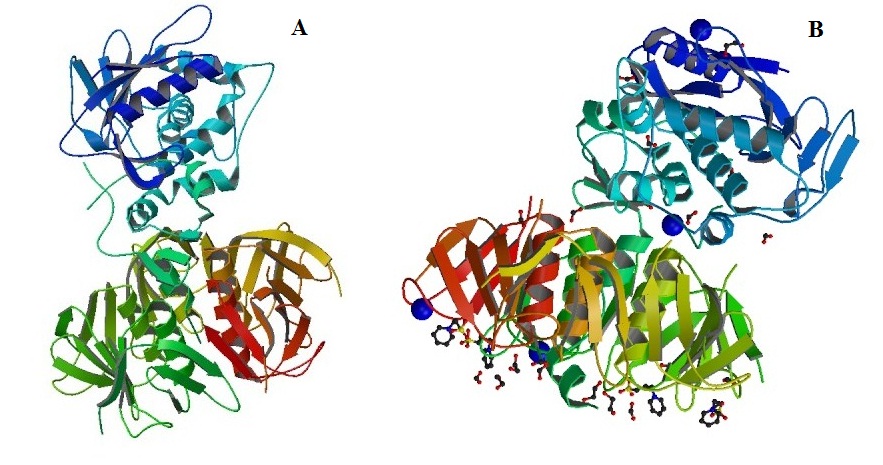}
\par\end{centering}

\caption{3-D structures of Stx1(A) and Stx2 (B), showing its differences. (Taken
from RCSB Protein Data Bank)}
\end{figure}

In eukaryotic cells the surface receptor for members of the Stx family
is the neutral glycosphingolipid globotriaosylceramide (Gb3; gal-$\alpha$ $\left[1\rightarrow4\right]$gal
$\beta$ $\left[1\rightarrow4\right]$Glc-ceramide), with exception
of Stxe variant that recognizes the globotetraosylceramide (Gb4; galNAc-$\beta$ $\left[1\rightarrow3\right]$gal
$\alpha$ $\left[1\rightarrow4\right]$gal $\beta$ $\left[1\rightarrow4\right]$Glc-ceramide).
Several synthetic analogues of Gb3 exists and these are able to distinguish
amongst Stx variants with a modification in the N-acetyl group. Complicating
this picture is the fact that many cellular binding sites to Stx are
available, and each of the subunits in the B pentamer have three Gb3
binding sites \cite{adrianabdifferentiation,okuda_targeted_2006}.

Gb3 is synthesized in the Golgi of eukaryotic cells and is transported
to the plasma membrane, having its trisaccharide residue out-membrane
and the non covalent ceramide hydrocarbon in the plasma membrane.
The joining subunit of Stx specifically recognizes the terminal alpha
1,4 trisaccharide digalactose. However, not only the part of carbohydrate
is important in the toxin joining mechanism, also the lipid tail is
important in the interaction of this toxin with its receptor. The
sensitivity of the cell to the toxin and the linking procedure between
the toxin and the cell can be regulated by many factors. For instance,
changes in the transduction signals can be important for such regulations:
an incubation prolonged with AMPc, or the exposition to butyric acid
can induce the production of new receptors for the Shiga toxin. Cytokines
(IL-1), tumoral necrosis factor (TNF) induced by Shiga toxin or LPS
during the infection time could induce Gb3 synthesis in many types
of cells. The induction of Gb3 production triggers the action of Shiga
toxins and causes severe complication during the infection. The internalization
involves formation of a clathrin-coated pit on the cell membrane.
In some cells, vesicles-bound toxins are subjected to the fusion with
the cellular lysosome, resulting in toxin degradation. In those cells
Stx-sensitives, the endosomal vesicle contain a toxin-complex receptor
which is transported to the endoplasmatic reticule. 

During this process the subunit A is cleaved by a protease of membrane
furin-like, generating a catalytically active 27 kD A1 N-terminal
fragment, and a 4 kD A2 C-terminal fragment which remains attached
by a disulphide bound. The active catalytically fragment for this
dimer exerts its effect upon the ribosomes when\textcolor{black}{{}
is released}\textcolor{red}{{} }in the cytoplasm \cite{paton_pathogenesis_1998,sandvig_shiga_2001,saito_structure-dependent_2012}.

Toxin can exerts its effect in eukaryotic cells by one of following
three mechanisms: firstly, the ribosomes inactivation and the inhibition
of synthesis of the cytoplasmic protein causing the cellular death.
Secondly, the depurination in the ribosomes generates a unique response
of signals translation named \textquotedblleft{}ribotoxic stress response
\textquotedblright{} (RSR), that leads to the production of cytokines,
or others factors that resulting in many different events including
the apoptosis of the cells implicated. In the last site, the joint
of the holotoxin Stx or its subunit at the receptor, can start a cascade
of cytoplasmic signals translation different of the response RSR.
The end result of this events is the cellular death (apoptosis or
necrosis) or an response of inflammation in the cells that remain
viable, and maybe others immediate responses \cite{obrig_escherichia_2010}.
\\

\section{Detection }

\subsection{Biological Methods}

It is considered that an appropriate response to bacterial outbreaks
consist in its early detection and the use of adequate antibiotics
to control them. The methods used to detect food pathogens have been
grouped in four mainly categories. In the first group are the conventional
microbiological methods, in which the food is mixed with selective
medium enriched to increase the population of a target organism. Also,
in this categories is located the agar plating in selective or differential
media to isolate the pure culture, and the test of the culture by
means of phenotype analysis or take of metabolic fingerprinting. These
conventional microbiological methods, which are considered the gold
standard, are reliable and accurate, but require much time and are
very labor intensive to obtain the results \cite{sulan_bai_rapid_2010}.
Typically, the traditional methods involve a series of steps: pre-enrichment,
selective enrichment, biochemical screening and serological confirmation.
These step are laborious, requiring significant amount of time, expensive
equipment and trained personal \cite{van_hattum_functional_2012}. 

The detection of microorganisms which produce Stx has been difficult,
due to the microorganism diversity and the detection limits of the
samples in the environment \cite{mauro_shiga_2011}. The bacterial
detection methods have to be rapid and very sensitive because, as
explained before, the presence of even a single pathogenic organism
in the body or food may become infectious. Extremely selective detection
methodology is required, due to the big numbers of non-pathogenic
organisms, which are often present in a complex biological environment,
and coexist with the pathogenic bacterias. The traditional methods
used for bacterial identification comprises the counting of cells
with the aid of a optical microscope or by flow cytometry; measuring
physical parameters by piezocrystals, impedimetry, redox reactions,
optical methods, calorimetry, ultrasound techniques and detecting
cellular compounds such as ATP, DNA, protein, lipid derivatives and
radioactive isotopes \cite{dmitri_ivnitski_biosensors_1999}. 

The second group is composed by variants of the polymerase chain reaction
(PCR) technique. Numerous articles has been published employing these
methods in the determination of pathogens amongst them, \textit{E.coli}
in food . The methods to detect and identified the verotoxin produced
by \textit{E.coli,} based in these PCR genetic techniques, have been
developed very recently. An advantage of this techniques is that with
small amounts of samples make possible the accurate detection of the
target species \cite{naravaneni_rapid_2005}. \textcolor{black}{The
PCR technique is very sensible for this task, but require hours to
process the sample. Besides considerable bio-molecular skills are
needed to prove the identification. The whole process could take from
48 hours to five days to obtain results, to this time is needed to
sum the time expended in sample transportation, from the controlled
site to the lab}\textcolor{blue}{{} }\cite{subramanian_rapid_2012}\textcolor{blue}{.
}Moreover, the presence of certain genes has not, necessarily, a direct
correlation with dangerous toxin levels \cite{van_hattum_functional_2012}. 

The PCR in real time is the most used to the quantification of specific
fragments of DNA. The amount of product synthesized during PCR is
measured in real time by the detection of a fluorescent signal produced
to result of a specific amplification. This methods is rapid and sensitive,
but in some cases false positive and negative results are obtained,
requiring an additional confirmation through to the hybridization
probe and polymorphism of length of restriction fragment \cite{sulan_bai_rapid_2010}.
Several methods based in real-time PCR have been commercialized for
Shiga toxin genes analysis in microbiologic food \cite{margot_evaluation_2013}. 

The third group includes the methods of immunosorbent assay (ELISA),
its principle is based in the join of an antibody to the target antigen.
It is a accurate and precise method, ideal for the quantitative and
qualitative detection of many types of protein in complex matrices,
when searching objectives are known. The sensitivity is low and up
to 3 or 4 hours are is necessary in order to complete the analysis
\cite{sulan_bai_rapid_2010}. The ELISA for Stx1 and Stx2 can be used
in stool sample. There are also commercial immunoassay for the detection
of the Shiga-like toxin in milk and chopped meat \cite{kehl_evaluation_1997,teel_rapid_2007}. 

\textcolor{black}{One of detection methods based in the capture of
these toxins with antibodies was realized by }\textcolor{black}{\emph{Hattum
et al}}\textcolor{blue}{\emph{.}} The presence of verotoxin in the
sample is verified through a fluorescence signal coming from an antibody-antibiotin
complex. The method is complementary to genetic methods, because it
allow the detection of mRNA but does not provides information about
if the toxin is only expressed or it is functional \cite{van_hattum_functional_2012}. 

\textit{\emph{The researcher }}\textit{Ashkenazi et al}\textit{\emph{,
have evaluated the efficacy of a Gb3-based ELISA for detection of
Shiga toxin, both from culture pure plates and, most importantly,
directly from a mixed bacterial culture}}\cite{ashkenazi_rapid_1989}.
The glycodendrimers and \textcolor{black}{glycol-conjugated nanoparticles}\textcolor{red}{{}
}has been used as anti-adhesive molecules for toxins and biosensors
to monitor the protein-carbohydrate interactions. The glyconanoparticles
show the best potential for studying these protein-carbohydrate interactions.
The\textcolor{green}{{} }\textcolor{black}{size distribution of these}\textcolor{green}{{}
}nanoparticles \textcolor{black}{is reasonably narrow} and is comparable
with the size of the studied biomolecules,\textcolor{blue}{{} }\textcolor{black}{moreover,
these structure have been well described and are easily manipulated
chemical structures.}\textcolor{red}{{} }The glycopolydiacetylene nanoparticle
(GPDA) has been used to monitor receptor-ligand events junctions for
viruses, toxins, bacteria and antibody-receptor interactions, due
to its unique colorimetric transition when joining to these macromolecules.
Therefore, due to the well-known affinity of Shiga toxin towards the
final {[}gal-$\alpha$1,4-gal{]} disaccharide unit, GPDA nanoparticles
containing this specific disaccharide sequence, has been used to identify\textit{
E.coli} Shiga toxin in in 96-well plates \cite{nagy_glycopolydiacetylene_2008}. 

The fourth, and most recent group of detection methodologies is based
on microarray techniques. These methods allow the simultaneous identification,
in foods, of a large number of pathogens with a simple reaction. The
basic idea is that many complementary probes are joined in \textcolor{black}{a}
matrix shape\textcolor{black}{{} on} a solid surface, each one of these
site contains \textcolor{black}{several }copies of specific probes.
The matrix is hybridized with the DNA isolated from the sample of
interest, producing a characteristic fluorescence. During this step
of hybridization, the fragment is joined to the probe about the bases
of DNA complementarity. However, this regular microarrays method needs
an expensive equipment, both for exploring that matrix and for compiling
the generated data \cite{sulan_bai_rapid_2010}.

\subsection{Detection by Biosensors}

Many of the reported methods for the detection of pathogenic bacteria
are applied to \textit{Escherichia coli}. Most \textit{E.coli}-specific
methods have a detection limit between $10{}^{3}$ and $10{}^{5}$
cells/mL and some rely on the amplification of specific genes of the
\textit{E. coli} genome for specific identification \cite{noguera_carbon_2011}.
Structure-based, designed synthetic biosensors have been recently
considered as attractive alternative to conventional platforms for
pathogens detection. Moreover, these biosensors have important characteristics,
such as high grade of sensitivity and detection specificity, minimum
effort in sample preparation, profitability, miniaturization and portability
to real-time monitoring while reduce the total time requested for
detection. Therefore, a great effort has been allocated the development
of these fast biosensors of diverse nature, as they are considered
promising devices for pathogenic bacteria detection \cite{wang_advances_2013}.

Ideal attributes of any element of recognition would be a great stability,
facility of immobilization in the sensor platform and the specificity
of recognition to the host with minimal cross-reactivity to other
pathogens agents \cite{singh_recent_2013}. In the detection of biochemical
and physiological processes, the biosensors have been converted in
a fundamental tool. 

According to the methods used for signal transduction, biosensors
are divided in four groups: mass, optical, electrochemical, and thermal
sensors. taking into account the way used for target identification,
biosensors can be classified into two categories: sensors for direct
detection and sensors with indirect detection. Direct detection biosensors
are designed in such a way that the bio-specific reaction is directly
determined in real time by measuring the physical changes induced
by the complex formation. Whereas indirect detection biosensors are
those in which a preliminary biochemical reaction takes place and
then, one of some products of this reaction are detected by a sensor. 

Very often, they have been grouped into the following categories:
biosensors based on direct detection of bacteria, flow-injection biosensors,
monitoring bacterial metabolism, detection of enzyme labels, genosensors
and the emerging artifical nose \cite{dmitri_ivnitski_biosensors_1999}.
The mot popular bio-probes employed for these surface biosensors for
pathogens detection are nucleic acid, antibody, entire phages, phage
display peptide (PDP) and more recently the\textcolor{black}{{} phage\textquoteright{}s
receptor binding proteins}. Biosensors does not suffer for a long
time of sample pre-enrichment, \textcolor{black}{nor a step of secondary
enrichment}\textcolor{green}{,} therefore, they can predict the level
and type of contamination of food. All this, faster than biological,
microbiological, immunological and conventional molecular methods
\cite{singh_recent_2013}. \\

For long time nucleic acid lateral flow immunoassays (NALFIA) has
been used as biosensors for the detection of nucleic acid. In these
assays, nucleic acids can be captured on the lateral flow test strips
by means of relation of independent or dependent antibodies. Amongst
the advantages of lateral flow strips can be mentioned: one-step,
simplicity, fast results, low cost, versatile and a prolonged shelf
life. Moreover if compared with traditional electrophoresis detection,
NALFIA has other advantages like shorter response time, and no needs
hazardous reagents. NALFIA can be used as biosensor for bacterial
detection due to its characteristics: test finished after 5 -10 min,
species selectivity, direct measurement, portable and designed for
field application. This is complemented with a simple manufacturing
process. Finally, if NALFIA strips are used with coloured nanoparticles,
an easy visual detection can be afforded. This combination overcomes
the need of have expensive equipments which is a major shortcomings
in the application of immunosensor techniques, a fact acknowledged
by \textit{Noguera et al,} whom have reported the use of a carbon
nanoparticles-NALFIAs combination for rapid detection and identification
of genes encoding various STEC virulence factors (vt1, vt2) \cite{noguera_carbon_2011}.

The technique of Localized Surface Plasmon Resonance (LSPR) which
is based in the analyzing intermolecular interactions of biomacromolecules
have been used for detection of biological toxins. In the recently
study developed by \textit{Nagatsuka et al}, was used a portable LSPR
detection system that uses glyco-chips containing Au nanoparticles
coated with synthetic oligosaccharides that specifically bind toxins.
The ricin, Shiga toxin, and cholera toxin were selected as the targets
in this study. For every case, the LSPR detection was completed within
20 min and was highly specific to the target toxin. Because of its
sensitivity, a LSPR system based on glyco-nanotechnology is competitive
with other techniques and has the added advantage of being portable
and of providing simple and rapid analysis in contaminated areas \cite{nagatsuka_localized_2013}.

In an assay of rapid detection, efficient, accurate and of low cost,\textit{
Bai et al,} have developed methods to detect pathogens transmitted
by food in the surface of and optical biosensor thin film. An advantage
of this technology is that, due to the characteristic optic of the
thin film to the biosensors chip, the experimental results can be
visualized by the human eyes without specific instruments. Therefore,
this technology avoids an initial investment on expensive instruments
and can distribute to any laboratory of individual research with basic
facilities for molecular biology. Generally, these methods are rapids
and robust, has an excellent sensitivity and specificity, and are
quite competitive, in terms of cost, when compared with existing technologies
\cite{sulan_bai_rapid_2010}.

\textit{Tu} \textit{et al} in 2006, have developed a biosensor based
on the immobilization of an antibody on a optic fiber for fast detection
of low levels of \textit{E.coli }O157: H7 and to Shiga-similar toxins
in ground beef sample. The principle of the sensor is an type sandwich
immunoassay using an antibody specific to\textit{ E.coli} O157: H7
or its toxins. A polyclonal antibody was first immobilized on polystyrene
fiber waves guides through a reaction of biotin-streptavidin, which
served as entity for capturing bacterias and toxins. The fluorescent
molecules immobilized on the fiber were excited by the evanescent
wave, and the portion of the emission light was transmitted by the
fiber and finally collected in the photo-detector at 670-710 nm \cite{tu_fiber-optic_2006}.

\textit{\emph{Another similar approach was used by}}\textit{ Ngundi
et al} in 2006, which have studied an array-based technique that provides
the capability to perform multiple analysis simultaneously. They report
a technique for immobilizing sugars onto planar waveguides and employing
the patterned arrays to analyze carbohydrate-binding protein toxins.
In this study are used an array biosensor, and are employing two monosaccharide-derivatives:
N-acetylneuraminic acid (Neu5Ac), and N-acetyl galactosamine (GalNAc)
as receptors for protein toxins \cite{ngundi_detection_2006}.

\section{Conclusion}

Large outbreaks of human illness with severe manifestations due to
STEC infections have occurred in recent decades caused by Shiga toxin
producers microorganisms. The high infectious capacity of these microorganism
make dangerous even the presence of few individuals in a given medium,
as a results these episodes has received great attention by the community
scientific. 

These outbreaks have taken place due to an historical little monitoring
or surveillance in the food supply chain. This incomplete control
on food or its precursors is attributable, in part, to the lack of
standardized methods for the detection or enumeration of these bacteria
in food matrices, but they are also due to the lack of consensus on
which serotypes are most important. Still, other issues difficult
the battle against pathogens, such as: the rapid urbanization, the
increasing number of poor and hungry people around the world, the
rapid development of transportation, environmental change and the
human activity. 

The war between humans and pathogens never ends. New diseases are
continuously appearing, and these organism will find more chances
to interact with humans. Therefore, we must take advantage of new
advances in technology and the strengthen the search to develop new
diagnostic mechanisms and design new drugs to fight against infectious
diseases.

In this scenario, it is a challenge to create techniques to detect
the bacterias involved in these outbreaks with high degree of efficiency.
In recent times, the use of structure-based designed biosensors have
emerged as an alternative, with the necessary properties for reliable
and effective use in routine applications. Many researchers are in
the search of biosensor systems with specificity to distinguish the
target bacteria in a multi-organism matrix, the sensitivity to detect
bacteria directly, the adaptability to detect different analytes without
sample pre-enrichment and the capacity to give real-time results.
At the same time, the biosensor must have relatively simple and inexpensive
configurations. In the case of STx, little have been done in this
field, no robust sensors for continuous water and food monitoring
have been developed. Therefore, the necessity to explore new ways
to develop robust, long-time of use sensors, exists, and its a challenge
to be accomplished in the next years

\bibliographystyle{unsrt}
\addcontentsline{toc}{section}{\refname}

\bibliography{bibliografia}

\end{document}